\documentclass{article}

\usepackage{arxiv}

\usepackage[utf8]{inputenc} 
\usepackage[T1]{fontenc}    
\usepackage{hyperref}       
\usepackage{url}            
\usepackage{booktabs}       
\usepackage{amsfonts}       
\usepackage{nicefrac}       
\usepackage{microtype}      
\usepackage{lipsum}		
\usepackage{graphicx}
\usepackage{natbib}
\usepackage{doi}
\usepackage{framed,multirow}

\usepackage{pifont}
\newcommand{\cxmark}{\ding{55}}
\usepackage{atbegshi}
\AtBeginDocument{\AtBeginShipoutNext{\AtBeginShipoutDiscard}}

\title{An efficient semi-supervised quality control system trained using physics-based MRI-artefact generators and adversarial training}

\author{Daniele Ravi \texttt{d.ravi@herts.ac.uk} \\
1) School of Physics, Engineering and Computer Science\\ University of Hertfordshire, Hatfield, UK \\
2) Centre for Medical Image Computing (CMIC) \\ Department of Computer Science \\University College London, UK\\
3) Queen Square Analytics, London, UK
\And for the Alzheimer's Disease \\Neuroimaging Initiative*
\And Frederik Barkhof \\ 1) Centre for Medical Image Computing (CMIC)\\ Department of Medical Physics and Biomedical Engineering\\ University College London, UK\\
2) Department of Radiology and Nuclear Medicine, Neuroscience Campus Amsterdam\\ VU University Medical Center, Amsterdam, the Netherlands\\
3) Queen Square Analytics, London, UK\\
4) NMR Unit, Queen Square Multiple Sclerosis Centre, Department of Neuroinflammation\\ Queen Square Institutes of Neurology, Faculty of Brain Sciences, University College London, London, UK\\
5) Department of Brain Repair and Rehabilitation,  Queen Square Institute of Neurology\\ University College London, London, UK\\
\And Daniel C. Alexander \\1) Centre for Medical Image Computing (CMIC) \\ Department of Computer Science \\ University College London, UK\\
2) Queen Square Analytics, London, UK
\And Lemuel Puglisi \\ Queen Square Analytics, London, UK\\
\And Geoffrey JM Parker \\1) Centre for Medical Image Computing (CMIC)\\ Department of Medical Physics and Biomedical Engineering \\ University College London, UK\\
2) Queen Square Analytics, London, UK\\
3) NMR Unit, Queen Square Multiple Sclerosis Centre\\ Department of Neuroinflammation\\ Queen Square Institutes of Neurology\\ Faculty of Brain Sciences\\ University College London, London, UK
\And Arman Eshaghi \\1) Centre for Medical Image Computing (CMIC)\\ Department of Computer Science, University College London, UK\\
2) Queen Square Analytics, London, UK\\
3) NMR Unit, Queen Square Multiple Sclerosis Centre\\ Department of Neuroinflammation\\ Queen Square Institutes of Neurology\\ Faculty of Brain Sciences\\ University College London, London, UK}

\renewcommand{\shorttitle}{Daniele Ravi \textit{et~al.}}

\begin{document}
\maketitle
\begin{abstract}
Large medical imaging data sets are becoming increasingly available. A common challenge in these data sets is to ensure that each sample meets minimum quality requirements devoid of significant artefacts. Despite a wide range of existing automatic methods having been developed to identify imperfections and artefacts in medical imaging, they mostly rely on data-hungry methods. In particular, the scarcity of artefact-containing scans available for training has been a major obstacle in the development and implementation of machine learning in clinical research.
To tackle this problem, we propose a novel framework having four main components: (1) a set of artefact generators inspired by magnetic resonance physics to corrupt brain MRI scans and augment a training dataset, (2) a set of abstract and engineered features to represent images compactly, (3) a feature selection process that depends on the class of artefact to improve classification performance, and (4) a set of Support Vector Machine (SVM) classifiers trained to identify artefacts. Our novel contributions are threefold: first, we use the novel physics-based artefact generators to generate synthetic brain MRI scans with controlled artefacts as a data augmentation technique. This will avoid the labour-intensive collection and labelling process of scans with rare artefacts. Second, we propose a large pool of abstract and engineered image features developed to identify 9 different artefacts for structural MRI. Finally, we use an artefact-based feature selection block that, for each class of artefacts, finds the set of features that provide the best classification performance. We performed validation experiments on a large data set of scans with artificially-generated artefacts, and in a multiple sclerosis clinical trial where real artefacts were identified by experts, showing that the proposed pipeline outperforms traditional methods. In particular, our data augmentation increases performance by up to 12.5 percentage points on the accuracy, F1, F2, precision and recall. At the same time, the computation cost of our pipeline remains low --less than a second to process a single scan-- with the potential for real-time deployment. Our artefact simulators obtained using adversarial learning enable the training of a quality control system for brain MRI that otherwise would have required a much larger number of scans in both supervised and unsupervised settings. We believe that systems for quality control will enable a wide range of high-throughput clinical applications based on the use of automatic image-processing pipelines.
\end{abstract}

*Data used in preparation of this article were obtained from the Alzheimer's Disease Neuroimaging Initiative (ADNI) database (adni.loni.usc.edu). As such, the investigators within the ADNI contributed to the design and implementation of ADNI and/or provided data but did not participate in analysis or writing of this report. A complete listing of ADNI investigators can be found at: \url{http://adni.loni.usc.edu/wp-content/uploads/how_to_apply/ADNI_Acknowledgement_List.pdf}


\section{Introduction}
Large and well-organized data sets are key for training and validating machine learning solutions. Real-world data sets are critical to enabling the development of robust approaches for clinical use. A feature of such data sets is that they contain data corruption, such as image artefacts or errors in data acquisition, that can degrade machine learning method performance. A quality control system to identify and remove corrupted samples is essential. Quality control and artefact removal are key for automatic image-analysis pipelines embedded in clinical workflow~\citep{saeed2022image} and large-scale data-collection initiatives. 

In general, when a substantial artefact appears in the target area of the image (e.g. inside the brain), the scan may not be suitable for use for diagnostic purposes, for research or by downstream algorithms to aid clinical decisions~\citep{hann2021deep}. Visual inspection by experts has traditionally been used to evaluate image quality and identify potentially problematic scans. However, this solution is time-consuming, does not scale to large amounts of data, suffers from relatively poor inter-rater reliability that is typical of human experts, has large overhead costs and is not suitable for real-time data streams. Automatic deep learning methods offer an alternative, but they may require substantial computational resources (i.e. use of GPUs) at training and inference time, as well as very large data sets. Finally, they often fail to generalise well across data centres. 

The development of an automatic system that detects artefacts in real-time and with minimal resource requirements (reduced training set, no GPUs, etc.) would bring significant benefits at little cost. Such a system would improve diagnosis efficiency by repeating problematic scans instantly instead of requiring repeat patient visits to the hospital, freeing time for the scanner that can be used to examine other patients. From a research point of view, a quality control system is useful to remove corrupted images from large datasets to increase the statistical power of a study.

As of today, automatic approaches to quality control can be divided into four classes of solutions that depend on the type of training (supervised and unsupervised) and the granularity of artefacts identified (pixel-based, image-based). In supervised learning, because samples with artefacts are not as frequent as the artefact-free samples, it is difficult to create a balanced dataset required for the training. Unsupervised approaches, instead try to first learn the artefact-free image distribution and then identify artefacts by finding samples out of the distribution. To do so they require large datasets of artefact-free images that adequately represent the entire population heterogeneity. These large datasets are not always available or are difficult to collect. Therefore, unsupervised approaches often fail to generalise and lack adequate fine-tuning (i.e. finding the optimal threshold to separate images with artefacts from the good ones). 

Generating new images to augment the training set can be an alternative solution for these training problems. However, since artefacts are scarce and originate from a wide range of root causes, state-of-the-art generative models, such as~\citep{schlegl2019f}, often learn only the distribution of normal images (artefact-free images) instead of focusing on the generation of artefacts.
In addition, it is very challenging to simulate MRI scans that look realistic and that, at the same time, are artefact-free.

To overcome the limitations above we propose to corrupt existing MRI scans to add controlled artefacts as an alternative solution to simulating new artefact-free MRI scans. Therefore, instead of learning the distribution of artefact-free images, we developed a set of generators that learn to create MRI with controlled artefacts in a data-efficient manner.

Since data with artefacts are very limited, fully data-driven generators are not easy to be trained. For this reason, our generators are based on MR physics domain knowledge and are obtained by using a set of parametric functions to create specific types of artefacts. The parameters control the severity of each artefact and they are learned by using adversarial training on a set of artefact-free images. This allows the generation of large and diverse data sets that otherwise would be unfeasible to collect in the real world.

After building the proposed artefact generators, we extract a combination of data-driven and engineered features from both corrupted and artefact-free scans to build a suitable image representation that could be used for the identification of artefacts inside the images. To extract these features, we use the traditional imaging domain and the k-space domain (the radio-frequency domain where the MRI scans are acquired). Currently, only a small number of approaches use the k-space domain to identify artefacts (e.g.~\citep{shaw2020heteroscedastic,stuchi2020frequency}). However, due to the nature of the problem, the k-space is essential to generate and identify MRI artefacts and therefore to assist in the training of an automatic quality evaluation model for MRI.

Finally, our proposed quality control models include a novel artefact-based feature selection block, developed to find the best set of features for each class of artefact. Selected features are then used to train a set of SVM classifiers and detect images with artefacts in real-time.
 
Our main objectives in this study are: i) to determine that brain scans with generated artefacts, obtained by physics-based artefact generators, can be used to augment an available training set and to improve the classification model, especially in comparison with unsupervised approaches based only on learning the artefact-free image distribution (i.e. ~\citep{schlegl2019f}) and ii) to combine a pool of brain imaging features that provides a robust and efficient solution to identify scans with artefacts in real-time.

\section{Related Work}
Below we review the state-of-the-art literature on quality checking of medical images. We have identified four classes of work.

\subsection{Supervised - Pixel-level classification} 
Supervised deep learning models have achieved impressive results in a wide range of medical applications. Supervised training requires a large amount of data, paired with precise labels, which are obtained by experts' evaluation and which introduce significant data preparation costs. Many supervised approaches have been proposed to identify artefacts and control the quality of images using segmentation~\citep{monereo2021quality}. They usually exploit a subset of hand-labelled segmented images obtained by experts who have delineated the artefacts at the pixel/voxel level. Supervised learning is then performed by deep neural networks, often based on a U-Net encoder-decoder architecture~\citep{ronneberger2015u}. Fully convolutional networks (CNNs) have also shown excellent results, even when trained on small datasets~\citep{ben2016fully}. More advanced work, such as~\citep{ali2021deep}, proposed to detect and classify artefacts based on a framework that combines a multi-scale convolutional neural network detector with an encoder-decoder model aimed to identify irregularly shaped artefacts. 

More recent works have also shown that attention-based supervision can be used to alleviate the requirement for a large training dataset for training~\citep{li2018tell}. For example,~\citep{venkataramanan2020attention} has proposed using attention maps as an additional supervision cue and enforcing the classifier to focus on all artefact-free regions in the image. 

\subsection{Supervised - Image-level classification}
This class of approaches aims to bypass the classification at the voxel level, thereby reducing the required computational cost. They are based on training supervised models that require both high and low-quality images and predict the quality scores using a set of scored images labelled by experts. For example,~\citep{bottani2021automatic} developed a supervised method based on CNN to compute quality scores. To train and validate the model, they asked trained raters to annotate the images following a visual pre-defined QC protocol. Similarly, in~\citep{ma2020diagnostic} they proposed to use several supervised CNN-based frameworks capable of assessing medical image quality and detecting if an image can be used for diagnostic purposes. In particular, they visualised activation maps from different classes to investigate discriminating image features learned by the model. In a similar vein,~\citep{graham2018supervised} conducted a study that introduced a CNN-based approach for reducing the reliance on manual labelling in a supervised setting. The approach utilized simulated data for training and a small amount of labelled data for calibration and was demonstrated to be effective in detecting severe movement artefacts in diffusion MRI.

\subsection{Unsupervised - Pixel-level classification}
Unsupervised methods are an alternative to supervised approaches. They often rely on training a generative model that learns the distribution of artefact-free images. Once trained, the model is used to identify potential artefacts by comparing an input image with the generated normal counterparts, and anomalies are identified by measuring the reconstruction error between the observed data and the model-generated image. The idea behind this is that these generative models, trained on only artefact-free images, cannot properly reconstruct anomalies. Approaches based on Generative Adversarial Networks (GANs)~\citep{baur2020steganomaly,schlegl2019f,schlegl2017unsupervised,sun2020adversarial} and Variational Auto-Encoders (VAEs)~\citep{chen2018unsupervised,you2019unsupervised,pawlowski2018unsupervised}, which use reconstruction error, have been widely employed in the literature. 

In a similar research direction,~\citep{an2015variational} introduced a method for artefact detection that relies on the reconstruction probability, which is defined as the likelihood that the decoded image matches the original input. This probability can be used to identify potential anomalies in the input data. Specifically, areas with low reconstruction probabilities are more likely to contain artefacts, while pixels with high reconstruction probabilities are more likely to represent the underlying signal. Therefore, by examining the distribution of reconstruction probabilities across the input data, one can identify regions that are likely to contain artefacts and focus further analysis on those regions.

Recent works in the field of unsupervised knowledge distillation and representation learning have also developed algorithms to identify abnormalities. For example, it is possible to train a set of small networks (called students) to replicate the exact behaviour of a larger network (called teacher), and abnormalities can be measured by computing the difference between the students and the teachers~\citep{bergmann2020uninformed}. If the outputs are different it means that the students fail to generalize and a possible anomaly is occurring. Additionally, the student networks' uncertainty can be used as a scoring function for anomalies. 
Another solution is proposed in~\citep{pinaya2022unsupervised}, which combines the latent representation of vector quantised variational autoencoders with an ensemble of autoregressive transformers to obtain an unsupervised anomaly detection and segmentation on FLAIR images from the UK Biobank dataset. The combination of these 2 frameworks was proposed to overcome the limitations of transformers which demand a very large dataset and high computational resources to have a good performance~\citep{trenta2022explainable}. 

Unsupervised approaches often identify the class thresholds required to separate the anomalies from the artefact-free images. Accurate thresholds are not easily identified. To tackle this problem~\citep{silva2021looking} proposed a novel formulation that does not require accessing images with artefacts to define these thresholds. In particular, they obtain this by an inequality constraint, implemented by extending a log-barrier method.

\begin{figure*}[t]
\begin{center}
\includegraphics[width=1\textwidth,trim={0cm 10cm 5cm 0cm},clip]{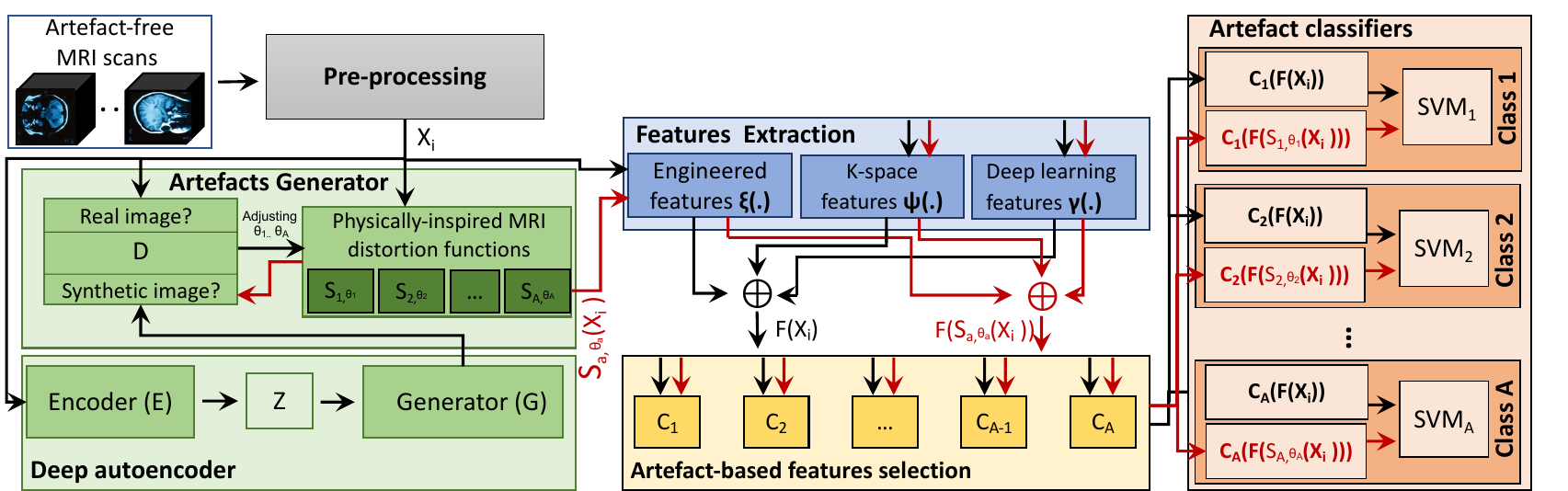}
\caption{Proposed pipeline developed to detect artefacts from MRI scans.\label{fig:pipeline}}
\end{center}
\end{figure*}

\subsection{Unsupervised - Image-level classification}
Similarly to the previous class, these approaches are trained in an unsupervised manner which requires only images without artefacts during training. Additionally, they are also developed for real-time processing obtained by working at the image level.
For example,~\citep{mortamet2009automatic} proposed an unsupervised approach developed for real-time processing where a set of quantitative tools are used to quickly determine artefacts in MRI volumes for large cohorts. These tools are based on fast quality control features developed to detect image degradation, including motion, blurring, ghosting, etc. Similarly,~\citep{sadri2020mrqy} proposed a framework that can identify MR images with variation in scanner-related features, field-of-view, image contrast, and artefacts by using a set of quality measures and metadata designed for real-time filtering. These measures can be used as a feature representation to fit a binary (accept/exclude) classifier and identify when abnormalities occur~\citep{esteban2017mriqc}.
Another unsupervised approach that has been proposed to find fast features is proposed in~\citep{oksuz2019automatic}. This approach automatically detects the presence of motion-related artefacts in cardiac MRI. In particular, it uses a 3D spatiotemporal CNN and a long-term recurrent convolutional network. However, since the data set is highly imbalanced -- a relatively small number of images with artefacts when compared with the number of good-quality images -- they propose a data augmentation technique to alter the k-space and generate realistic synthetic artefacts. Following a similar direction,~\citep{shaw2020heteroscedastic} integrates four different simulated classes of artefacts that can be used for extending existing training data. Our artefact generators are inspired by these ideas while bringing novelty with more classes of artefacts, as outlined in Table~\ref{table:simulator_paramters}.

\subsection{Limitations of the state-of-the-art}
The use of supervised learning is not always straightforward. Obtaining training labels is time-consuming, expensive and subject to mistakes typical of human raters and their bias. Additionally, since pure artefacts are very rare, collecting a sufficiently large dataset of these labelled artefacts may not be feasible.

Several problems limit the use of unsupervised approaches as well. First, it is hard to collect a large set of medical images where no artefacts exist and where the data represent the entire population heterogeneity. The lack of a representative dataset with millions of images makes it infeasible to learn when an abnormality comes from an actual artefact or unseen patients with unusual structures. Second, in an unsupervised setting, where no abnormal samples are available for training, it is hard to identify the class thresholds required to separate the anomalies from the artefact-free images.

One of the main disadvantages of the approaches working at the pixel level is that they often do not take into consideration computational constraints (i.e real-time processing) and they perform computationally expensive operations, which require powerful GPUs to process an entire 3D MRI efficiently. Additionally, although voxel classifications or segmentation produce fine details, these solutions are often not suitable for modalities with lower-resolution scans.

The approaches working at the image level provide greater benefits since they offer a time-efficient solution obtained by bypassing the classification at the voxel level. However, these approaches still require a large dataset of artefact-free images for unsupervised learning, where the model learns to identify artefacts without any prior knowledge of their presence. For supervised learning, where the model is trained to classify images with and without artefacts, a combination of artefact-free and artefact-containing images is necessary to ensure the model can accurately distinguish between the two.

We chose to implement a system that belongs to this last category, but in contrast with the existing approaches, we propose a semi-supervised adversarial training strategy aiming to generate artefacts obtained by physics-based MRI-artefact generators and able to augment the available training set. In particular, we avoid using images with artefacts during training (which can be rare) and our generators are trained to add artefacts by finding the minimum level of corruption for which images are not considered anymore artefact-free. A combination of features selected for each class of artefact (extracted from the artefact-free images and the corrupted ones obtained from our generators) are then used to train a set of supervised SVM artefact classifiers.
To the best of our knowledge, we are the first to build a system to augment a training dataset by using artefact generators trained using adversarial learning and aimed to find the best parameters that describe the severity of the artefacts.

\section{Methods}
Our proposed pipeline is depicted in Fig.~\ref{fig:pipeline} and consists of five blocks coded with different colours: 
i) a pre-processing block, ii) a set of artefact generators that take pre-processed artefact-free images as input and generate images with controlled artefacts, iii) a features extraction block, iv) a feature selection process, and v) an ensemble of SVM classifiers.

Our pipeline generates artefacts throughout the entire 3D volume by iterating the artefact simulation on each individual 2D brain slice. To corrupt the entire 3D volume consistently, we create artefacts with the same severity and parameters on each slice extracted from the first view (i.e., axial). We believe this approach adequately generates artefacts in 3D, which can be more challenging to achieve. Directly modelling artefacts in 3D would depend on the specific artefact considered. For the majority of the artefacts we considered (i.e., noise, smoothing, bias, banding, Gibbs, folding, and zipper), we do not see a significant difference in modelling them in 2D versus 3D. However, for some artefacts such as motion, directly modelling it in 3D would be more realistic. We chose to focus on the 2D plane to simplify the image generation process, which would have otherwise required a significant amount of effort.  

To train our system we only use a subset of these 2D slices. In particular, in our training, we use a data set of $X_n$ with $n=1,2,...N$ representing artefact-free brain $T_{1}$-weighted MRI scans from which we extract $x_{k,v,n}\in R^{s \times s}$ pre-processed slices from $K=3$ fixed positions and $V=3$ different views.

We provide the details of each block of our pipeline in the following sections.

\subsection{Pre-processing}
The pre-processing block aims to reduce irrelevant variations in the data and prepare each input MRI $X_n$ to train the model. To allow real-time performance, we exclude computationally intensive pre-processing operations often used in medical imaging, such as non-linear image registration. 

Our pre-processing consists of five steps: i) removing 5\% of intensity outliers from the entire MRI volume (this essential step ensures that the successive intensity standardisation can act optimally); ii) performing slice-wise intensity standardisation (zero mean, unit standard deviation), iii) auto-cropping each slice based on OTSU threshold~\citep{xu2011characteristic}; iv) normalising the intensity values in the range [-1,1], and reshaping each slice to a fixed size of $s \times s$ with $s$=300. In cases where the original resolution is smaller than our designated fixed resolution (a scenario that encompasses the majority of instances within our dataset), we apply zero-padding to the images. This process ensures that all images conform to our predetermined resolution without compromising their content or quality. It's crucial to emphasize that the proposed resolution aligns with the typical requirements for 1mm MRI scans, which are the images found in the ADNI dataset. Therefore, we recommend using images with a resolution of 1mm in conjunction with our pipeline to achieve results similar to those obtained in our study.
The recommendations for using a resolution of 300x300 pixels have been also validated through consultations with two expert neurologists, each boasting more than a decade of specialized knowledge and experience in this field. Their expert assessment confirmed the suitability of this resolution for this particular study.

Regarding the process of eliminating intensity outliers this is accomplished by identifying and then removing exceptionally high or low-intensity values within the MRI data. This is achieved by establishing a threshold based on the 95th percentile of intensity values and subsequently clipping all values that surpass this threshold. These intensity outliers can arise from various factors, including noise or inconsistencies in the MRI scanner. Our primary goal in this procedure is to eliminate potential artefacts, thus obtaining a class of artefact-free images which can be utilized for training and validation purposes. The application of a 5\% threshold (95th percentile) to remove intensity outliers is consistent with established practices in the literature~\citep{hadjidemetriou2009restoration}.

\subsection{Physics-based artefacts generation}\label{sec:simulator}
The next step in our pipeline is to build a set of $A$ generative models $S_{a,\theta_a}$ (one per each class of artefact $a$) having parameters $\theta_a$. Each model is a degradation/corrupting function designed to create a class of artefacts occurring during the MRI acquisition. 

To build our artefact generators, we study the cause of different brain MRI artefacts and we emulate them by corrupting artefact-free images accordingly. For example, a low-pass filter applied in k-space would simulate a blurring artefact, while the addition of random spikes in k-space creates a banding artefact~\citep{moratal2008k,heiland2008aliasing}. 

\subsubsection{A proof-of-principle framework to generate artefacts}
\begin{figure*}[t]
\begin{center}
\includegraphics[width=0.9\textwidth,trim={0cm 1cm 10cm 0cm},clip]{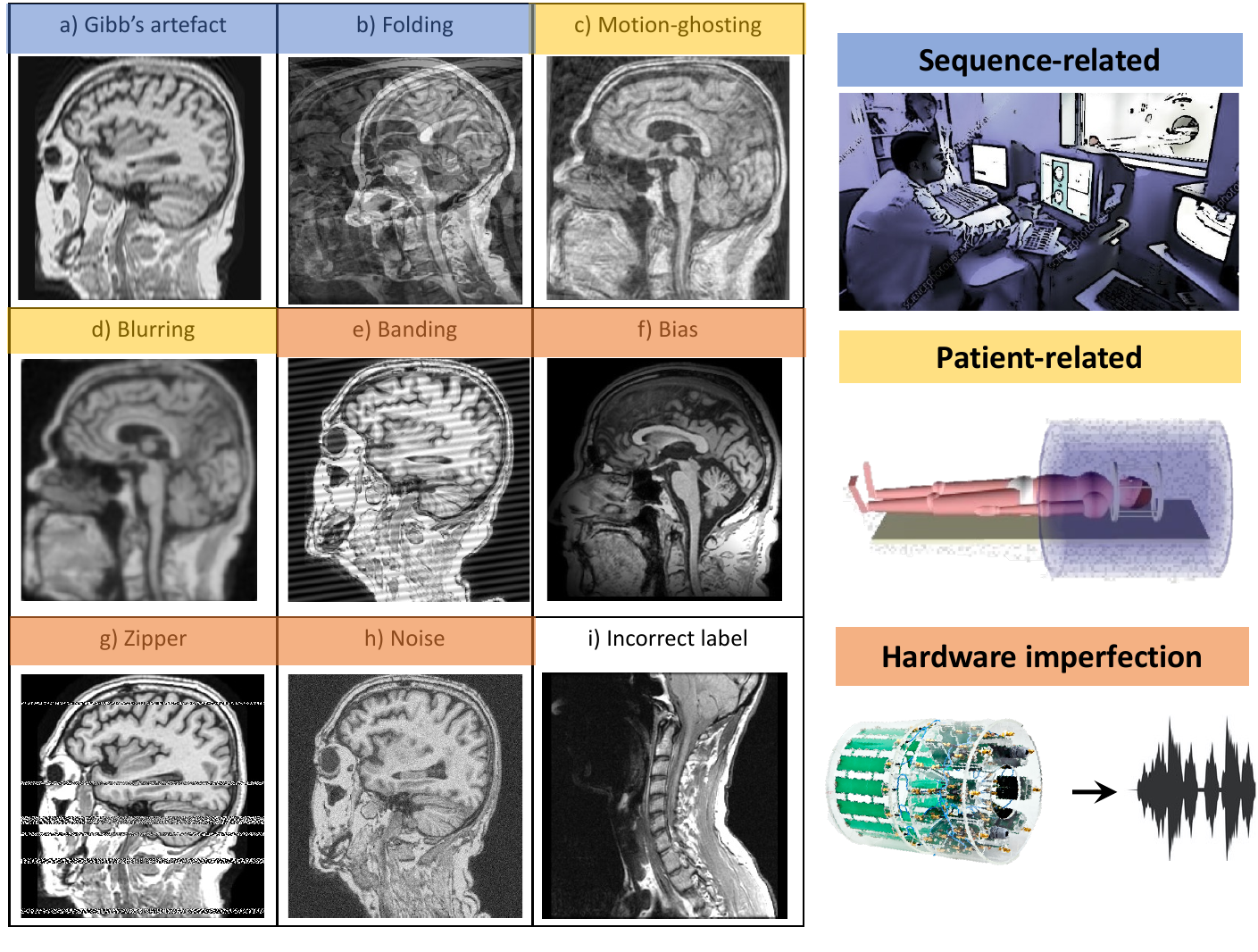}
\caption{Extreme cases of artefacts randomly generated by our artefact generators and organized in three different classes. In blue sequence-related artefacts, in yellow patient-related artefacts and in orange hardware-related artefacts.\label{fig:artefacts}}
\end{center}
\end{figure*}

The artefacts that we are considering in this study are not intended to be exhaustive but serve as a proof of concept to demonstrate the power of our solution, which can feasibly be extended to new artefacts for MRI and new modalities. In particular, we have identified 9 different common artefacts for $T_{1}$-weighted brain MRI divided into three groups: (1) hardware imperfection artefacts (i.e. noise from measurements, nonuniformity in the static magnetic field or nonuniformity in the radiofrequency field; (2) patient-related artefacts (e.g. ghosting and other motion artefacts); (3) sequence-related artefacts (e.g. Gibb's artefact, folding and blurring).  
We have also added a further category to account for mislabeled images. This category pertains to cases where the MRI scan does not fully show the brain or shows it only partially, but is still labelled as a brain MRI (for instance, when an image of the spinal cord is mistakenly labelled as a brain MRI). We consider this a potential source of error since it can have a detrimental impact on the performance of automated image analysis algorithms. A summary of all these artefacts is shown in Fig.~\ref{fig:artefacts} and summarised in Table~\ref{table:simulator_paramters}.

The implementation details of each of these artefacts are provided in the following sub-sections.

\begin{center}
\begin{table*}[t]
\centering
\caption{Summary of Artifacts, Related Parameters, and Corresponding Ranges considered to generate them.}\label{table:simulator_paramters}
\begin{tabular}{l|l|l}
\textbf{Artifact} & \textbf{Parameter}                                         & \textbf{Range}     \\ \hline
Folding           & Spacing size added to the K-space (in lines)                                      & $[1, 6]$           \\ \hline
Ghosting            & Rotation angle                                             & $[0, \pi/2]$       \\ \hline
Ghosting            & Degree of translation                                      & $[0, w/20]$, $[0, h/20]$  \\ \hline
Ghosting            &  Percentage of K-space lines to swap                                         & $[1\%, 50\%]$        \\ \hline
Gibb's             & Percentage of K-space lines to remove (vertically) & $[1\%, 30\%]$     \\ \hline
Gibb's             & Percentage of K-space lines to remove (horizontally) & $[1\%, 30\%]$     \\ \hline
Bands             & Amplitude of the spike (Intensity)                                          & [Mean Intensity, Max Intensity]            \\ \hline
Bands             & Distance from the center                                       & $[\frac14 h, \frac34 h]$, $[\frac14 w, \frac34 w]$\\ \hline
Bands             & Number of Corrupted points                                    & $[1, 8]$ \\ \hline
Blurring         & Standard deviation $\sigma$                                & $[0.8, 6]$       \\ \hline
Zipper            & Number of zipper regions                                          & $[1, 8]$            \\ \hline
Zipper            & Artifact size (in lines)                                   & $[1, 21]$          \\ \hline
Noise             & Standard deviation $\sigma$                              & $[0.05, 25.0]$       \\ \hline
Bias field        & Polynomial degree                                          & $[2,7]$            \\ \hline
Bias field        & Coefficients range lower bound                             & $[0.2, 1.2]$       \\ \hline
Bias field        & Coefficients range upper bound                             & $[0.7, 2.7]$       \\
\end{tabular}
\end{table*}
\end{center}

\subsubsection{Gibb's artefacts}
Truncation or Gibb's artefacts appear as a ringing effect associated with sharp edges at transitions between tissues of differing signal intensity (Fig.~\ref{fig:artefacts} a). This artefact is due to i) Fourier transforms reconstruction obtained from a finite sampled signal and ii) a lowering of the sampling coverage in k-space used to speed up the acquisition process. In our framework, we have implemented Gibb's artefacts by under-sampling k-space in both the frequency and phase encoding direction (see Section~\ref{k-space} for more details on k-space). In particular, to reduce the sampling coverage from high-quality images already acquired, we exclude the most peripheral information of k-space during Fourier reconstruction.

Two parameters control the severity of this artefact: the amount of data (number of lines or columns) removed from the k-space in each of the frequency and phase encoding directions.
\subsubsection{Folding artefacts}
Folding, or wrap-around, artefact corresponds to the spatial mismapping, or overlapping, of structures on the opposite side of the image from where they may be expected (Fig.~\ref{fig:artefacts} b). These artefacts are caused by corruptions occurring during the spatial encoding of objects outside the selected field of view. These can overlap the information inside the field of view. To emulate this artefact we follow the work proposed in~\citep{moratal2008k}, which increases the spacing between phase-encoding lines, thereby emulating a rectangular field of view, which creates the wrap-around effect.

The parameter that controls the severity of this artefact is the spacing size added between the lines of k-space.

\subsubsection{Patient motion: Ghosting and blurring}
MRI scan time is usually relatively long in order to generate high-resolution images. Therefore, motion artefacts are often unavoidable and are one of the most frequent issues during an MRI scan. The most frequent motion artefacts are: ghosting (Fig.~\ref{fig:artefacts} c), and blurring  (Fig.~\ref{fig:artefacts} d). We emulate blurred images by applying a Gaussian low-pass filter in k-space. Ghosting is emulated by aggregating two k-space matrices, generated from two slightly different versions of the same images. The first is an input image, the second is the same input image where a random affine transformation is applied to emulate the desired patient's motion. 

We adjust the severity of the ghosting artefacts using 3 parameters: the degree of the rotation and translation for the affine transformation, and the amount of k-space data that is taken from the second image and replaced with the k-space from the first image.

We adjust the blurring artefact by increasing the size of the Gaussian low-pass filter used.
\begin{center}
\begin{table*}[t]
\caption{Summary of existing and proposed features extracted from the imaging domain and used to represent the scans during the artefact detection task.}\label{table:existing_features}
\scalebox{0.95}{
\begin{tabular}{l|c}
 \textbf{Existing Engineered Features -- Imaging Domain} & \textbf{Reference}\\ 
 \hline
Mean, range, and variance of the foreground intensities &~\citep{sadri2020mrqy}\\
Coefficient of variation of the foreground &~\citep{wang2019fully} \\
Contrast per voxel&~\citep{chang2015reference}\\
Peak signal-to-noise ratio of the foreground &~\citep{sage2003teaching}\\
Foreground standard deviation divided by background standard deviation &~\citep{bushberg2011essential}\\
Mean of the foreground patch divided by background standard deviation &~\citep{esteban2017mriqc}\\
Foreground patch standard deviation divided by the centred foreground patch standard deviation &~\citep{sadri2020mrqy}\\
Mean of the foreground patch divided by mean of the background patch &~\citep{sadri2020mrqy}\\
Contrast to noise ratio &~\citep{bushberg2011essential}\\
Coefficient of variation of the foreground patch &~\citep{sadri2020mrqy}\\
Coefficient of joint variation between the foreground and background &~\citep{hui2010fast}\\
Entropy focus criterion &~\citep{esteban2017mriqc}\\
Foreground-background energy ratio &~\citep{shehzad2015preprocessed} \\
Global contrast factor on the background &~\citep{matkovic2005global}\\
Global contrast factor on the foreground &~\citep{matkovic2005global}\\
 \hline
 \textbf{ Proposed Engineered Features -- Imaging Domain} &\\ 
 \hline
Max and variance on the edge detector response obtained on the foreground& Proposed\\
Mean, variance and Shannon entropy on the edge detector response obtained on the background& Proposed\\
Min and max value of the integral over the row on the foreground& Proposed\\ 
Min and max value of the integral over the column on the foreground& Proposed \\

\end{tabular}
}
\end{table*}

\end{center}

\subsubsection{Band artefacts}
Gradients applied at a very high duty cycle, or other electronic interference can produce spikes in k-space~\citep{moratal2008k}. These spikes result in banding artefacts visible in the reconstructed image (Fig.~\ref{fig:artefacts} e). The location of these spikes in the k-space determines the angulation and the band pattern that affect the image.

We emulate these artefacts by corrupting a small number of points in a k-space line by adding a very high-intensity value compared with the rest of the k-space. The parameters that control this artefact are the amplitude of the spike, the maximum distance from the centre of k-spaced where the spike can happen, and the number of points corrupted. 

\subsubsection{Bias artefacts}
Bias field or intensity inhomogeneity is caused by spatial variations in the sensitivity of the acquisition coil and/or by spatial variation in the transmitted RF field. Generally, such intensity variations occur at a low spatial frequency across the image (Fig.~\ref{fig:artefacts} f). Although robust approaches for correcting these artefacts are today available (N3 and N4 bias field correction~\citep{tustison2010n4itk,boyes2008intensity}), we decided to include this artefact in our study because in severe cases and some high field settings (e.g. 7-tesla MRI), this problem still requires some attention. 
Following the work proposed in~\citep{van1999automated} we emulate the bias field as a linear combination of polynomial basis functions. We control the severity of this by using three parameters that represent the degree of the considered linear functions.
\subsubsection{Zipper artefacts}
Zipper artefact is generated by radio frequency interference. This can happen for example when a device or equipment (e.g. a mobile phone) is left in the scanning room during the acquisition. The result of this problem is an abnormal black-and-white signal band across the entire image, which we emulate by adding random lines of black-and-white pixels on the reconstructed image (Fig.~\ref{fig:artefacts} g). The parameters that control the severity of these artefacts are the number of zipper regions that can occur in an image and the max size of each of these regions.

\subsubsection{Noise artefacts}
Components of MRI scanners (e.g. coils, electronic components, etc.), electronic interference in the receiver system, and radio-frequency emissions due to the thermal motion of the ions in the patient's body can lead to noise in the final images (Fig.~\ref{fig:artefacts} h). We chose to model the noise in k-space by adding an error with a zero-mean Gaussian distribution to each of the acquired k-space samples.
The parameter used to control the severity of this artefact is the amplitude of the error (sigma of the Gaussian) used to perturb the raw data.

\begin{figure*}[t]
\begin{center}
\includegraphics[width=1\textwidth,trim={0cm 10cm 0cm 0cm},clip]{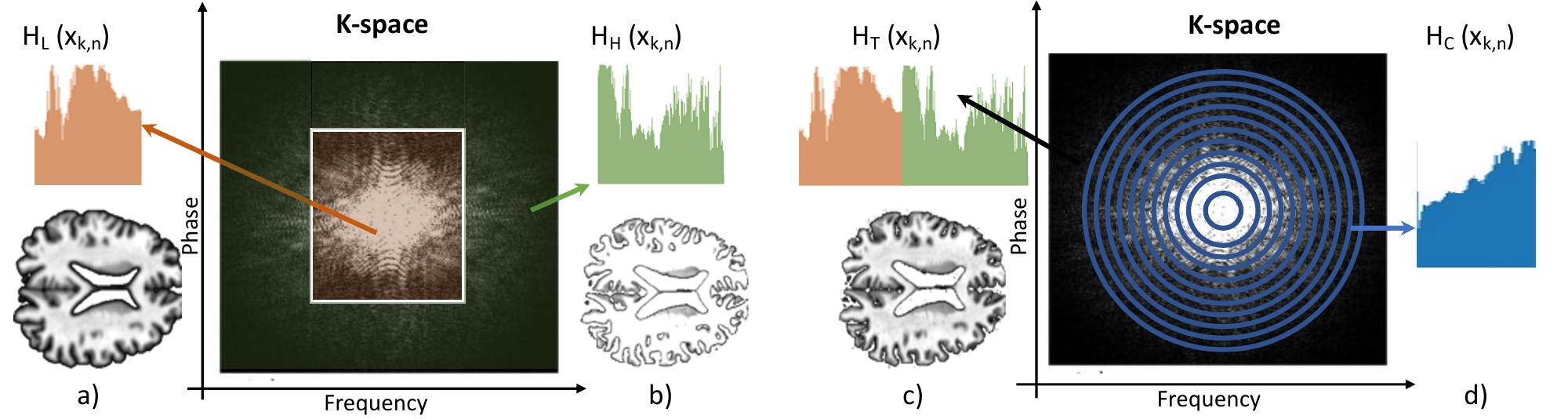}
\caption{The figure shows four different histograms ($H_L$, $H_H$, $H_T$ and $H_C$) computed from four areas of the k-space and used to extract statistical features from each scan.\label{fig:k-space-features}}
\end{center}
\end{figure*}

\subsection{Finding the optimal parameters: Adversarial training}
All the parameters $\theta_a$ controlling each artefact generator $S_{a,\theta_a}$ with $a \in $ ${1..A}$ are summarized in Table~\ref{table:simulator_paramters}. As described in the sections above, these parameters are developed to control directly the severity of each class of artefact. For example, when $|\theta_a|$ is close to zero the severity of the artefacts of class $a$ is small and the artefacts in the generated images are barely visible. At the same time, if $|\theta_a|$ is high, the minimum severity for this class of artefacts will be high, and strong artefacts are always generated. Indeed different values of $\theta_a$ will create different types of images used for augmenting the training set and this affects the final classifiers. In particular, if the artefacts are too small the classifier may not be able to learn how to separate artefacts from the normal images, whilst when the artefacts are too strong, the classifier may learn to separate them well in artefact-simulated images, but it will not generalise to real data.

To find the optimal parameters for each $S_{a,\theta_a}$ we exploit adversarial training which is a technique used in generative adversarial networks (GANs) to learn the distribution of a target training set. However, since we do not have a target training set of artefacts that we could use to learn this distribution, we exploit adversarial training to find only the minimum level of corruption for which images are not considered anymore artefact-free. Regarding the distribution of these artefacts, we instead assume that this will be uniform across the range of severity starting from the identified minimum value. We believe that this assumption will not hamper the training of our classifiers. In fact, this simply means that since we are not able to model directly the distribution of artefact, we train the classifiers also with extreme cases distributed uniformly that are instead unlikely to occur in the real world. 

In particular, for our adversarial training, we make use of a convolutional generative adversarial network (DCGAN)~\citep{radford2015unsupervised}. DCGAN has a discriminator $D$ that is trained adversarially with another network $G$ (the generator) via unsupervised learning. $D$ aims to discriminate realistic images from fake ones while $G$ is trained to fool $D$, i.e., to generate brain MRI with a similar distribution to the initial true distribution.

Our hypothesis is that once the DCGAN has been trained, the discriminator $D$ will acquire the ability to capture the natural variations present in images (without artefacts) and can serve as an evaluator to detect the appearance of synthetic artefacts in such images. This would enable us to learn the parameters that govern the degree of severity of these artefacts.

The loss functions used to train $D$ and $G$ are as follows:
{\begin{equation}\label{dbrain_loss}
L_{GAN}=\min\limits_{G}\max\limits_{D}\mathbb{E}_{k,v,n}\big[\log D\big(x_{k,v,n})\big) \big] + 
\mathbb{E}_{k,v,n}\big[1-\log D\big(G(z))\big)\big] ,
\end{equation}} 
where $\mathbb{E}$ is the expectation, $D$ estimates the probability that a slice belongs to the real distribution (i.e. artefact-free images) and $z$ is a latent vector obtained from $x_{k,v,n}$. Additionally, as we can see from Fig.~\ref{fig:pipeline}, $G$ takes as input the vector $z$ which is obtained by another encode network $E$ aimed to map the image domain to a latent space $z=E(x_{k,v,n})$ while $G$ acts as a decoder by mapping $z$ back to the image space. 

While $G$ and $D$ are trained simultaneously following a standard GAN schema (explained above), the networks $E$ and $G$ are trained following a convolutional autoencoder architecture using the loss

\begin{equation}\label{L_2}
L_{AE}= \mathbb{E}_{k,v,n}\frac{1}{s^2} || x_{k,v,n}- G(E(x_{k,v,n})) ||^2 ,
\end{equation}

where $||...||^2$ is the sum of squared pixel-wise residuals of values and $s^2$ is the number of pixels in the image. 

The use of three networks ($E$, $D$, and $G$) in our approach enables us to work directly with input images rather than random vector noise, as is typical in traditional GANs. The encoder ($E$) plays a critical role in this process by projecting the input image into a lower-dimensional representation, which enables the generator network ($G$) to produce high-quality synthetic images. Specifically, the encoder's task is to learn a representation that is relevant for generating realistic-looking data while ensuring that the generated data is indistinguishable from real data by the discriminator. By operating directly on input images and using the encoder to project them into a lower-dimensional space, our approach leverages the rich information already present in the input data to generate new, high-quality synthetic images.

Once $D$, $G$ and $E$ are set, we use $D$ to find the optimal parameters $\theta_a$ for our artefact generators. To do so, we minimize a new loss in Eq.~\ref{L_3}, which combines the euclidean 1-norm of $\theta_a$ and the discriminative loss obtained by using $D$ on generated images: 
\begin{equation}\label{L_3}
L_{S_a}= \mathbb{E}_{k,v,n}\big[ D\big(S_{a,\theta_a}(X_n)[k,v])\big] + ||\theta_a||_1 .
\end{equation}
An intuition behind our new formulation is that the second term of Eq.~\ref{L_3} ($||\theta_a||_1$) aims to decrease the amplitude of $\theta_a$ and minimise the artefacts. However, when $\theta_a$ becomes too small the images will not have realistic visible artefacts and consequently, $D$ cannot discriminate artefacts-free images from those with artefacts. We avoid this by controlling the discrimination loss $\mathbb{E}_{k,v,n}\big[ D\big(S_{a,\theta_a}(X_n)[k,v])\big]$, which will be high when $D$ is not able to discriminate the two classes. During the optimisation of Eq.~\ref{L_3} the parameters of the network $D$ are not trained but we use $D$ only to find the minimum values of $\theta_a$ for which the discriminator loss remains limited.

Once each $\theta_a$ is found we use our artefact generators to create new images to augment our training set.

\subsection{Feature extraction}\label{sec:features}
This block aims to extract a pool of efficient features (our image representation) from each slice, which will be used for the final classification of the scan. As we can see from Fig.~\ref{fig:pipeline} (section in blue), this block operates both on the real images $X_n$ and on the generated synthetic images $S_{a,\theta_a}(X_n)$. More specifically, for each normalised slice $x_{k,v,n}$ extracted from the MRI $X_n$ and for each $s_{a,k,v,n}=S_{a,\theta_a}(X_n)[k,v]$ representing the corrupted slices obtained by $S_{a,\theta_a}$ using the scan $X_n$, we extract three classes of features: i) engineered features $\xi(.)$ extracted from the imaging domain, ii) statistical features $\psi(.)$ extracted from the k-space domain, and iii) abstract features $\gamma(.)$ extracted using two popular deep neural networks.

All these features are extracted from every 2D slice. However, to ensure we capture 3D information in a computationally efficient manner, we implement a multiple slices configuration (2.5D, that is, 2D slices encompassing axial, sagittal and coronal views), where multiple slices are used at the same time.

In particular, in our 2.5D implementation, the features extracted from the slices at different view $v$ and the different position $k$ are all concatenated in a unique representation vector. The reason for this concatenation is that artefacts may appear in only a local area of the MRI volume and combining different views and different slices make it more likely to have at least one slice with visible artefacts in the proposed image representation. Therefore, for a generic MRI $X_n$ (with or without artefacts), our full set of features is

\begin{equation}\label{full_list}
F(X_n)=\prod_{i=1}^{K}\prod_{j=1}^{V}\oplus\Bigg[\xi(x_{i,j,n}),\psi(x_{i,j,n}),\gamma(x_{i,j,n})\Bigg] ,
\end{equation}

where $\oplus$ performs this concatenation between features' vectors. 

\subsubsection{Engineered features: Imaging Domain}
The first set of features that we propose are engineered features $\xi(.)$ developed to detect specific patterns in the imaging domain or to measure specific imaging characteristics such as first and second-order statistics (e.g., mean, variance, skewness, and kurtosis), signal-to-noise ratio, contrast per pixel, entropy-focus criterion, and ratios of different regions. All existing engineered features used in our pipeline are listed in the first part of Table~\ref{table:existing_features}.

Additionally, we propose nine new descriptors to identify local artefacts (e.g. zipper) and measure spatial consistencies inside each slice, which are not covered by previous existing methods. To do so, we exploit an edge detector based on a Laplacian filter and an image integral, obtained by integrating all intensity values along the row and column of a single slice. These new features are described in the second part of Table~\ref{table:existing_features}. In total, for each slice (imaging domain), we extract 26 engineering features.

\subsubsection{Statistical features: K-space}\label{k-space}
Signal in k-space represents spatial frequencies in the x and y directions rather than an intensity value describing a pixel value as in the imaging domain. In particular, each point ($k_x$,$k_y$) in k-space does not correspond to a single pixel (x,y) in the counterpart imaging domain, instead, they contain spatial frequency and phase information about every pixel in the reconstructed image. To reconstruct the MRI scan an inverse Fourier Transform is used to convert the k-space samples to the actual imaging intensities. 

We note that, in contrast to the imaging domain, correlations between consecutive points in k-space are less common, which is a characteristic often exploited by CNNs. Therefore, standard CNNs may not be ideal for use in k-space. Instead, to process such domain information we propose a set of statistical features. These features are identified as $\zeta(.)$ and are as follows: mean, standard deviation, skewness, kurtosis, interquartile range, entropy, coefficient of variation, k-statistic and an unbiased estimator of the variance of the k-statistic. 
We compute $\zeta(.)$ from four samples distributions obtained from different areas of k-space: i) the centre of k-space containing low spatial frequency information ($H_L$), ii) the peripheral area of k-space containing high-frequency information ($H_H$), iii) the entire k-space ($H_T$), and iv) an area obtained by integrating k-space samples within an annulus of signal between circles starting from the centre ($H_C$). In Fig.~\ref{fig:k-space-features} we show how each of these k-space areas is selected.
In total our k-space features consist of a vector $\psi(.)$ of 9 $\times$ 4 = 36 features formally defined as: 

\begin{equation}\label{k-space_features}
\psi(x_{k,v,n})= [\zeta(H_L(x_{k,v,n})),\zeta(H_H(x_{k,v,n})),\zeta(H_T(x_{k,v,n})),\zeta(H_C(x_{k,v,n}))] .
\end{equation}

\subsubsection{Deep learning features: Imaging Domain}
The last set of features $\gamma(.)$ are obtained in a fully data-driven fashion by using two popular deep learning networks: i) ResNet-101~\citep{he2016deep} pre-trained with IMAGE-NET where we use the last layer as a feature vector and ii) and a fast anomaly GAN network (f-AnoGAN)~\citep{schlegl2019f} where we use the reconstruction errors as a feature set. Since the last layer of ResNet-101 is very large (2048 nodes), we compress the obtained vector in a smaller representation using Principal Component Analysis (PCA). We keep the first 64 components that represent the highest explained variance (95\% of the variance): 

\begin{equation}\label{Deep_Learning_features}
\gamma(x_{k,v,n})= [PCA_{64}(ResNet(x_{k,v,n}),GAN(x_{k,v,n})].
\end{equation}

\begin{figure*}[t]
\begin{center}
\includegraphics[width=0.9\textwidth,trim={0cm 3cm 10.5cm 0cm},clip]{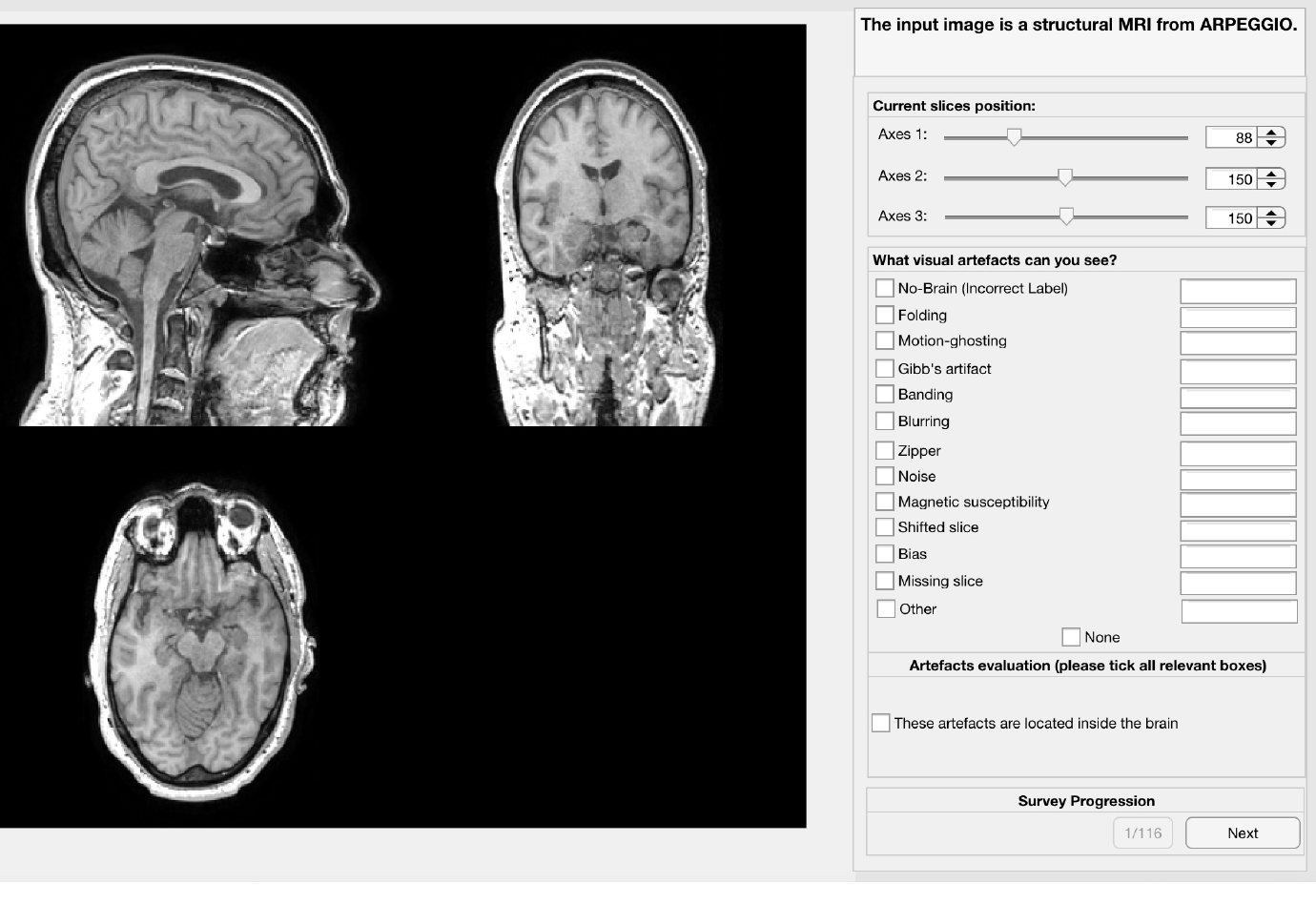}
\caption{Graphical user interface used to annotate images on real clinical data.\label{fig:survey}}
\end{center}
\end{figure*} 
\subsection{Artefact-based feature selection}\label{sec:selection}
In summary, our feature extraction block generates a feature vector that we identify as $F$. As mentioned above, some features could provide contrasting information when they operate on different classes of artefacts. For example, a feature that measures the sharpness of an image could be useful to identify blurring artefacts but it would not contribute when dealing with band artefacts or noise artefacts since it provides a high-value score for them. For this reason, we group features in different sets and measure their different performances to identify which combination provides the highest accuracy for each class of artefacts. 

Specifically, our artefact-based selection block will select for each class of the artefacts $a$ the best combination $c_a \in ( \xi, \psi, \gamma, \xi \oplus \gamma,\xi \oplus \psi,\gamma \oplus \psi, \xi \oplus \psi \oplus \gamma )$ so that the classification accuracy from the classifier $Q_a$ on the subset of samples from class $a$ is maximized,

\begin{equation}\label{features_selection_formula}
\max\limits_{c_a}\mathbb{E}_{k,v,n}\big[\log Q_a\big(c_a(F(x_{k,v,n}))\big) \big] +\mathbb{E}_{k,v,n}\big[1-\log Q_a\big(c_a(F(s_{a,k,v,n}))\big)\big] .
\end{equation}

To summarize, our feature selection method entails identifying the optimal combination of features from three distinct sets: imaging ($\xi$), k-space ($\psi$), and deep learning ($\gamma$). This selection process is performed for each type of artefact examined in our study.

\subsection {Ensemble of classifiers}\label{sec:classifiers}
In this section, we will define in more detail the classifiers $Q_a$ mentioned in the previous section. Since our feature vector is small we believe that the SVM model~\citep{hearst1998support} is a good solution for our problem and it can provide real-time performance even on a single processor. SVM is a powerful and fast classifier that constructs N-dimensional hyper-planes to optimally separate the data into two categories. The position of the hyperplane is determined using a learning algorithm that is based on the principle of structural risk minimization. Each $Q_a$ is trained using a different combination of features $c_a$ and optimized to discriminate only 2 classes of images: the images with artefacts $a$ from the artefact-free images. Formally, to train each $Q_a$ we make use of 2 sets of data points: i) $c_a(F(I_{i}),-1$, and ii) $c_a\big(F(S_{a,\theta_a}(I_{i}),+1)\big)$ both $\in [R^{d_a}, {\pm1}]$ and for $i \in {1..N}$. In this formulation, ${d_a}$ is the dimension of the considered feature set and the ${\pm1}$ is the binary class describing the label (artefact/no artefact).  
SVM takes these data points and outputs the hyperplanes that separate each of the two-class samples so that the distance between them is as large as possible~\citep{hearst1998support}. In our implementation, we use SVMs with radial basis function kernel.

To summarize, each classifier $Q_a$ is trained to identify a particular artifact. As each image is processed by all of the $Q_a$ classifiers, it is possible for an image to be classified as having one or more artefacts. However, if an image contains an unknown artefact, our system will assign it to the closest known ones or label it as "no artefact" if none of them is close enough, potentially causing a misclassification.

\begin{table*}[!ht]
\caption{Ablation study: Accuracy, F1, F2, Precision and Recall expressed in percentage and obtained by classifying the scans from $ADNI\_synt$ (synthetic images) with different configurations of the proposed pipeline.
In particular, $F$ represents the proposed feature extraction block, $C$ is the proposed feature selection, and $S$ is the data augmentation. We also evaluate two different versions: a 2D version where only the slices from the centre of each volume are used and a 2.5D where nine selected slices per volume are used.}\label{table:ablation_configuraion}

\begin{center}
\begin{tabular}{c||c|c|c|c|c}

Configuration&Accuracy (\%)&	F1(\%)&	F2(\%)&	Precision(\%)&	Recall(\%)	\\
\hline
$F\_2D$&	87.22 $\pm$ 13.24&	88.75 $\pm$ 11.96&	88.81 $\pm$ 11.00&	88.64 $\pm$ 12.64&	88.86 $\pm$ 10.42		\\
							
$F\_2.5D$&	84.55 $\pm$ 15.37&	85.71 $\pm$ 12.91&	86.52 $\pm$ 11.87&	84.40 $\pm$ 16.08&	87.07 $\pm$ 10.38		\\
							
$F\_C\_2D$&	92.59 $\pm$ 6.36&	92.30 $\pm$ 6.75&	92.21 $\pm$ 6.92&	92.45 $\pm$ 5.47&	92.16 $\pm$ 6.03		\\
							
$F\_C\_2.5D$&	91.80 $\pm$ 4.66&	91.67 $\pm$ 6.71&	91.18 $\pm$ 7.20&	92.49 $\pm$ 4.78&	90.86 $\pm$ 6.51		\\
							
$F\_S\_2D$&	96.69 $\pm$ 3.39&	96.75 $\pm$ 3.32&	96.72 $\pm$ 3.42&	96.80 $\pm$ 4.24&	96.70 $\pm$ 4.44		\\
							
$F\_S\_2.5D$&	97.88 $\pm$ 2.14&	96.87 $\pm$ 4.14&	96.90 $\pm$ 4.20&	\textbf{96.83 $\pm$ 3.10}&	96.91 $\pm$ 3.25		\\
							
$F\_C\_S\_2D$&	96.82 $\pm$ 3.06&	96.82 $\pm$ 4.05&	96.81 $\pm$ 3.05&	\textbf{96.83 $\pm$ 4.05}&	96.81 $\pm$ 4.05		\\
							
$F\_C\_S\_2.5D$&	\textbf{97.91 $\pm$ 2.08}&	\textbf{96.90 $\pm$ 3.08}&	\textbf{96.96 $\pm$ 3.04}&	96.80 $\pm$ 3.16&	\textbf{97.00 $\pm$ 3.00}		\\

\end{tabular}
\end{center}
\end{table*}

\begin{table*}[!ht]
\caption{Ablation study: Accuracy, F1, F2, Precision and Recall expressed in percentage and obtained by classifying the scans from $ADNI\_synt$ (synthetic images) with different numbers of slices and different views.}\label{table:ablation_views}

\begin{center}
\begin{tabular}{c||c|c|c|c|c}
Configuration&	Accuracy&	F1&	F2&	Precision&	Recall \\
\hline	
Axial&	96.82 $\pm$ 3.06&	96.82 $\pm$ 4.05&	96.81 $\pm$ 3.05&	96.83 $\pm$ 4.05&	96.81 $\pm$ 4.05\\	
						
Coronal&	95.36 $\pm$ 2.86&	95.12 $\pm$ 3.85&	95.33 $\pm$ 2.89&	94.78 $\pm$ 3.61&	95.47 $\pm$ 3.65\\	
						
Sagittal&	95.91 $\pm$ 2.63&	95.54 $\pm$ 3.85&	95.52 $\pm$ 2.88&	95.57 $\pm$ 3.75&	95.51 $\pm$ 3.72\\	
						
3-Axial&	97.86 $\pm$ 2.43&	96.61 $\pm$ 4.03&	96.49 $\pm$ 2.15&	\textbf{96.83 $\pm$ 3.87}&	96.40 $\pm$ 3.07\\	
						
3-Coronal&	97.60 $\pm$ 2.73&	96.55 $\pm$ 3.24&	96.59 $\pm$ 2.45&	96.50 $\pm$ 3.34&	96.61 $\pm$ 3.59\\	
						
3-Sagittal&	97.61 $\pm$ 2.45&	96.36 $\pm$ 3.06&	96.26 $\pm$ 2.67&	96.54 $\pm$ 3.80&	96.19 $\pm$ 3.45\\	
						
9 Slices&	\textbf{97.91 $\pm$ 2.08}&	\textbf{96.90 $\pm$ 3.08}&	\textbf{96.96 $\pm$ 3.04}&	96.80 $\pm$ 3.16&	\textbf{97.00 $\pm$ 3.00}\\	
\end{tabular}
\end{center}
\end{table*}

\begin{table*}[!ht]
\caption{Ablation study: Accuracy of our system in classifying scans from the $ADNI\_synt$ (Synthetic) dataset using a combination of three different feature sets: Imaging ($\xi$), K-Space ($\psi$), and Deep Features ($\gamma$). We evaluate the ability of each feature set to recognize different types of artefacts.}\label{table:ablation_featurees}
\begin{center}
\begin{tabular}{c||c|c|c|c|c|c|c|c|c}
Features &Incorrect-Label	&Folding	&Gosting	&Gibb's	&Banding	&Blurring	&Zipper&	Noise&	Bias \\
\hline	
$\gamma$&	97.42&	93.03&	96.23&	93.63&	92.44&	98.91&	97.67&	97.48&	91.13\\
$\xi$&	95.63&	91.66&	95.52&	93.49&	93.03&	98.56&	\textbf{98.53}&	97.15&	91.92\\
$\psi$&	92.67&	\textbf{98.03}&	97.00&	\textbf{98.66}&	92.55&	98.63&	97.59&	97.76&	91.24\\
$\xi$+$\psi$&	94.94&	97.10&	97.55&	98.28&	\textbf{94.98}&	98.08&	98.34&	97.61&	94.14\\
$\gamma$+$\xi$&	99.18&	93.27&	97.66&	93.61&	92.79&	97.00&	98.25&	\textbf{98.18}&	94.40\\
$\gamma$+$\psi$&	98.23&	97.35&	97.71&	98.44&	92.22&	99.06&	97.68&	97.61&	\textbf{94.80}\\
$\gamma$+$\xi$+$\psi$&	\textbf{99.68}&	97.96&	\textbf{98.77}&	98.31&	94.43&	\textbf{99.17}&	98.51&	97.79&	94.73\\
\end{tabular}
\end{center}
\end{table*}

\begin{table*}[!ht]
\caption{Quantitative comparison study: Accuracy, F1, F2, Precision and Recall expressed in percentage and obtained by comparing our approach against state-of-the-art methods on the test set from $ADNI\_synt$ (synthetic images). $F$ refers to the use of the proposed feature extraction method, $\mathbf{S_b}$ indicates the use of a one-class SVM, while $\mathbf{S_s}$ indicates the use of a two-class SVM trained using our corrupted images. The column named 'Augm.' indicates whether simulated artefacts were used during training or not.}
\label{table:comparison}

\begin{center}
\resizebox{0.98\linewidth}{!}{
\begin{tabular}{c|c|c||c|c|c|c|c}
\multicolumn{3}{c||}{Approach}&Accuracy (\%)&	F1(\%)&	F2(\%)&	Precision(\%)&	Recall(\%)	\\
Features&Classifier&Augm.&&&&&	\\

\hline
\textbf{F}&PCA-based&\cxmark&	68.79 $\pm$ 12.29&	67.69 $\pm$ 10.39&	67.62 $\pm$ 10.38&	67.80 $\pm$ 12.78&	67.58 $\pm$ 11.82		\\
							
\textbf{F}&Autoencoder&\cxmark&	78.66 $\pm$ 11.80&	78.19 $\pm$ 9.43&	77.87 $\pm$ 8.17&	78.73 $\pm$ 12.30&	77.66 $\pm$ 8.25		\\
							
\textbf{F}&\cite{an2015variational}&\cxmark&	82.30 $\pm$ 11.24&	80.04 $\pm$ 11.64&	79.36 $\pm$ 11.84&	81.19 $\pm$ 11.59&	78.92 $\pm$ 12.04		\\
							
\textbf{F}&\cite{zenati2018adversarially}&\cxmark&	79.91 $\pm$ 12.33&	80.69 $\pm$ 10.37&	80.69 $\pm$ 9.82&	80.70 $\pm$ 12.30&	80.69 $\pm$ 10.11		\\
							
\cite{schlegl2019f}&$\mathbf{S_b}$&\cxmark&	77.03 $\pm$ 14.07&	73.42 $\pm$ 23.10&	71.62 $\pm$ 23.48&	76.63 $\pm$ 14.97&	70.47 $\pm$ 24.35		\\
							
\cite{sadri2020mrqy}&$\mathbf{S_b}$&\cxmark &	81.61 $\pm$ 11.95&	79.20 $\pm$ 12.76&	77.35 $\pm$ 13.80&	82.50 $\pm$ 11.85&	76.16 $\pm$ 14.59		\\
							
\cite{schlegl2019f}&$\mathbf{S_s}$&\checkmark &	90.16 $\pm$ 15.24&	88.08 $\pm$ 26.54&	87.30 $\pm$ 26.59&	89.43 $\pm$ 14.49&	86.78 $\pm$ 26.56		\\
							
\cite{sadri2020mrqy}&$\mathbf{S_s}$&\checkmark &	94.67 $\pm$ 4.01&	94.58 $\pm$ 4.13&	94.43 $\pm$ 4.67&	94.82 $\pm$ 3.26&	94.34 $\pm$ 5.03		\\

\cite{szegedy2016rethinking}&\cite{szegedy2016rethinking}&\checkmark &94.23 $\pm$ 3.92&	94.76 $\pm$ 4.08&	94.41 $\pm$ 4.38&	95.35 $\pm$ 3.03&	94.18 $\pm$ 4.78\\
Proposed&Proposed&\checkmark &	\textbf{97.91 $\pm$ 2.08}&	\textbf{96.90 $\pm$ 3.08}&	\textbf{96.96 $\pm$ 3.04}&	\textbf{96.80 $\pm$ 3.16}&	\textbf{97.00 $\pm$ 3.00}		\\

\end{tabular}
}
\end{center}
\end{table*}

\begin{table*}[!ht]
\caption{Quantitative comparison study: Accuracy, F1, F2, Precision and Recall expressed in percentage and obtained by comparing our approach against state-of-the-art methods on the test set from the clinical trial. $F$ refers to the use of the proposed feature extraction method, $\mathbf{S_b}$ indicates the use of a one-class SVM, while $\mathbf{S_s}$ indicates the use of a two-class SVM trained using our corrupted images. The column named 'Augm.' indicates whether simulated artefacts were used during training or not.}
\label{table:comparisonARPEGGIO}

\begin{center}
\resizebox{0.98\linewidth}{!}{
\begin{tabular}{c|c|c||c|c|c|c|c}
\multicolumn{3}{c||}{Approach}&Accuracy (\%)&	F1(\%)&	F2(\%)&	Precision(\%)&	Recall(\%)	\\
Features&Classifier&Augm.&&&&&	\\
\hline
\textbf{F}&PCA-based&\cxmark&	83.44 $\pm$ 29.29&	90.08 $\pm$ 35.34&	86.35 $\pm$ 35.63&	97.06 $\pm$ 3.03&	84.03 $\pm$ 35.84		\\
							
\textbf{F}&Autoencoder&\cxmark&	83.33 $\pm$ 13.65&	90.35 $\pm$ 10.84&	87.31 $\pm$ 13.59&	95.92 $\pm$ 5.55&	85.39 $\pm$ 15.21		\\
							
\textbf{F}&\cite{an2015variational}&\cxmark&	84.28 $\pm$ 16.25&	90.61 $\pm$ 14.83&	87.65 $\pm$ 17.37&	96.01 $\pm$ 5.07&	85.78 $\pm$ 18.60		\\
							
\textbf{F}&\cite{zenati2018adversarially}&\cxmark&	86.36 $\pm$ 17.98&	91.84 $\pm$ 14.86&	88.90 $\pm$ 18.77&	\textbf{97.18 $\pm$ 2.47}&	87.05 $\pm$ 20.72		\\
							
\citep{schlegl2019f}&$\mathbf{S_b}$&\cxmark&	87.17 $\pm$ 11.30&	93.00 $\pm$ 6.73&	91.22 $\pm$ 8.80&	96.13 $\pm$ 4.19&	90.07 $\pm$ 10.14		\\
							
\citep{sadri2020mrqy}&$\mathbf{S_b}$&\cxmark&	83.62 $\pm$ 22.70&	86.00 $\pm$ 25.87&	84.76 $\pm$ 26.11&	88.14 $\pm$ 25.96&	83.96 $\pm$ 26.35		\\
							
\citep{schlegl2019f}&$\mathbf{S_s}$&\checkmark &	87.57 $\pm$ 12.94&	93.61 $\pm$ 8.45&	92.47 $\pm$ 12.14&	95.58 $\pm$ 5.15&	91.72 $\pm$ 14.33		\\
							
\citep{sadri2020mrqy}&$\mathbf{S_s}$&\checkmark &	92.20 $\pm$ 5.29&	96.00 $\pm$ 2.91&	96.30 $\pm$ 3.31&	95.51 $\pm$ 5.35&	96.49 $\pm$ 4.33		\\

\cite{szegedy2016rethinking}&\cite{szegedy2016rethinking}&\checkmark &	92.43 $\pm$ 5.19&	94.89 $\pm$ 4.22&	94.94 $\pm$ 4.52&	94.79 $\pm$ 3.56&	94.98 $\pm$ 4.09		\\					
Proposed&Proposed&\checkmark &	\textbf{94.76 $\pm$ 5.36}&	\textbf{96.37 $\pm$ 2.89}&	\textbf{96.99 $\pm$ 1.81}&	95.34 $\pm$ 5.49&	\textbf{97.42 $\pm$ 2.02		}\\
\end{tabular}
}
\end{center}
\end{table*}

\section{Dataset and Training Details}\label{sec:training}
Data used in the preparation of this article were obtained from the ADNI database (adni.loni.usc.edu). ADNI was launched in 2003 as a public-private partnership, led by Principal Investigator Michael W. Weiner, MD. The primary goal of ADNI has been to test whether serial MRI, positron emission tomography (PET), other biological markers, and clinical and neuropsychological assessment can be combined to measure the progression of mild cognitive impairment (MCI) and early Alzheimer's Disease.

In our experiments, we make use of two datasets. In the first dataset called $ADNI\_synt$, we asked three medical imaging experts to select n=4000 $T_{1}$-weighted MRI scans from the ADNI1 and ADNI2 datasets. The three experts (one radiologist and two medical imaging researchers) make sure to exclude scans with artefacts through a consensus procedure. On top of these scans, we add 36,000 synthetic MR images with artefacts, generated by our artefact generators $S_{a,\theta_a}$. We generate the same number of images for each class of artefact. 

Since the scans in ADNI are acquired using an MP-RAGE (Magnetization Prepared - Rapid Gradient Echo) sequence, the synthetic images that we create have the same type of contrast. Real scans from ADNI are labelled as artefact-free images and generated scans from our artefact generators are labelled as images with artefacts. We divided our dataset into a training set (Artefact-free MRI: 2000; Artefacts MRI: 2000x9), a validation set (Artefact-free MRI: 1000; Artefacts MRI: 1000x9), and a test set (Artefact-free MRI: 1000; Artefacts MRI: 1000x9). Each set is distinct, with unique images selected from ADNI, and diverse artefacts generated for each scan.

\begin{table*}[!ht]
\caption{Computation time required to process a single scan and obtained by our approach against existing real-time solutions. We record the computation time obtained for both feature extraction and final classification. $F$ refers to the use of the proposed feature extraction method, $\mathbf{S_b}$ indicates the use of a one-class SVM, while $\mathbf{S_s}$ indicates the use of a two-class SVM trained using our corrupted images.}
\label{table:computation_time}

\begin{center}
\resizebox{0.95\linewidth}{!}{
\begin{tabular}{c|c||c|c|c|c}

\multicolumn{2}{c||}{Approach} &Features Extraction (s)&	Classification ($10^{-5}$s)&	 Total Time (s)\\
Features&Classifier&&&&\\
\hline
\textbf{F}&PCA-based&0.8134 &7.033&0.8135\\
\textbf{F}&Autoencoder&0.8134 &11.176&0.8136\\
\textbf{F}&\citep{an2015variational}&0.8134 &5.489&0.8135	\\			
\textbf{F}&\citep{zenati2018adversarially}&0.8134 &2.613&0.8135\\
\citep{schlegl2019f}&$\mathbf{S_s}$&	\textbf{0.2731} & \textbf{0.127}& \textbf{0.2732}	\\
\citep{sadri2020mrqy}&$\mathbf{S_s}$& 0.4489 & \textbf{0.127} & 0.4490	\\
Proposed&Proposed&	0.8134 & 1.524 & 0.8136\\

\end{tabular}
}
\end{center}
\end{table*}

Although this dataset has a large sample size, some of the scans are generated by a synthetic process and therefore the results may not be representative of a real-world scenario. For this reason, we use an external testing dataset from a randomized clinical trial that had enrolled primary progressive multiple sclerosis (MS) patients~\citep{barkhof2015arpeggio} including i) 48 manually selected scans with expert-identified artefacts (3 with folding, 7 with motion, 8 with Gibbs' artefacts, 13 with blurring, 15 with noise and 2 with bias),  and ii) 48 randomly selected scans without artefacts. Using the GUI in Fig.~\ref{fig:survey} we asked three radiologists to label these 96 images according to the available 9 classes of artefacts. 

Each scan can be associated with multiple artefacts and can have two possible levels of severity (minor and major). All participants were instructed to use a common labelling protocol and the results were merged using a majority voting system.

To train our system we use a workstation provided with an 8-core CPU (Intel Xeon Bronze 3106 CPU @ 1.70GHz) and an NVIDIA GTX TITAN-X GPU card with 12 GB of memory on the training set from $ADNI\_synt$. To optimize the performance of our pipeline and find the best configuration for the feature selection block, we utilize the validation set. We also employ a random grid search technique using the validation set to tune hyperparameters such as the learning rate=1e-4 and batch size=64 for the networks. Once we identify the optimal hyperparameters, we apply them to the test set to obtain the final performance metrics for our pipeline. In particular, we utilized the $ADNI\_synt$ test set to evaluate the effectiveness of our proposed approach through an ablation study and a comparison with other state-of-the-art solutions. Additionally, we validate the generalizability of our solution by testing it on images from the MS clinical trial, which involves a different disease diagnosis than the training data. This last experiment helps us evaluate the performance of our system in a real-world scenario and provides more accurate results. It also indicates whether our artefact generators create realistic artefacts adequately.

During our adversarial training, we carefully monitored the convergence of the generator and discriminator. Initially, the generator produced low-quality data that the discriminator could easily distinguish from the real data, leading to a high loss for the generator and a low loss for the discriminator. However, as the training progressed, we observed a steady decrease in the generator's loss, indicating that it was learning to generate increasingly realistic data. Simultaneously, the discriminator's loss slightly increased until it reached a plateau, indicating that the generator was making its task more difficult.
Although we encountered some minor oscillations in the generator's loss during training, the system convergence after 500 epochs.

In our experiments, we realized that some of the existing methods used as a comparison approach are not developed for the full classification of artefacts and they are either developed to extract only features (without performing the final classification) or developed to perform only the classification (without the initial feature extraction).
Therefore to be able to compare our approach against theirs, we complement these approaches with the missing part (features extraction or classification) taken from our pipeline. Without this additional step, it would not be possible to analyze some of the existing approaches specifically on the task of artefact detection for brain MRI.

The blocks of our pipeline that we use to run the existing approaches are: i) $F$ -- the proposed features extraction (Section~\ref{sec:features}), ii) $\mathbf{S_b}$ -- the SVM classifiers (Section~\ref{sec:classifiers}), trained without data-augmentation -- one-class SVM and iii) $\mathbf{S_s}$ -- the SVM classifiers, trained using a dataset augmented with our corrupted images (Section~\ref{sec:simulator}) -- two-class SVM. In particular, the state-of-the-art approaches that we have considered are: i) PCA-based, ii) Autoencoder, iii) Variational Autoencoder~\citep{an2015variational} and iv)~\citep{zenati2018adversarially} and they were trained using the bloc $F$ that uses the features we have proposed in our pipeline. Additionally, we have considered two more unsupervised methods (~\citep{schlegl2019f} and~\citep{sadri2020mrqy}) designed to extract features, which we combined with our two classifier setups (blocks $\mathbf{S_b}$ and $\mathbf{S_s}$).
Finally, we compared our approach with a standard fully supervised Inception network trained on both original and simulated data~\cite{szegedy2016rethinking}.

While both feature extraction techniques ($F$) and data augmentation methods ($\mathbf{S_s}$) have the potential to enhance the performance of state-of-the-art approaches, their implementation may be constrained by various factors, such as the choice of features and the nature of the training approach. For example, the approaches proposed in \citep{schlegl2019f} and \citep{sadri2020mrqy} only used one class of features, which limited their ability to explore other feature combinations. On the other hand, unsupervised techniques such as PCA-based, Autoencoder, Variational Autoencoder \citep{an2015variational} and \citep{zenati2018adversarially} were trained exclusively on artefact-free images, and incorporating artefacts into these unsupervised training methods would require modifying the training process. As a result, we have refrained from applying data augmentation techniques ($\mathbf{S_s}$) to these unsupervised techniques.

\begin{figure*}[t!]
\begin{center}
\includegraphics[width=0.9\textwidth,trim={0cm 6cm 7.5cm 0cm},clip]{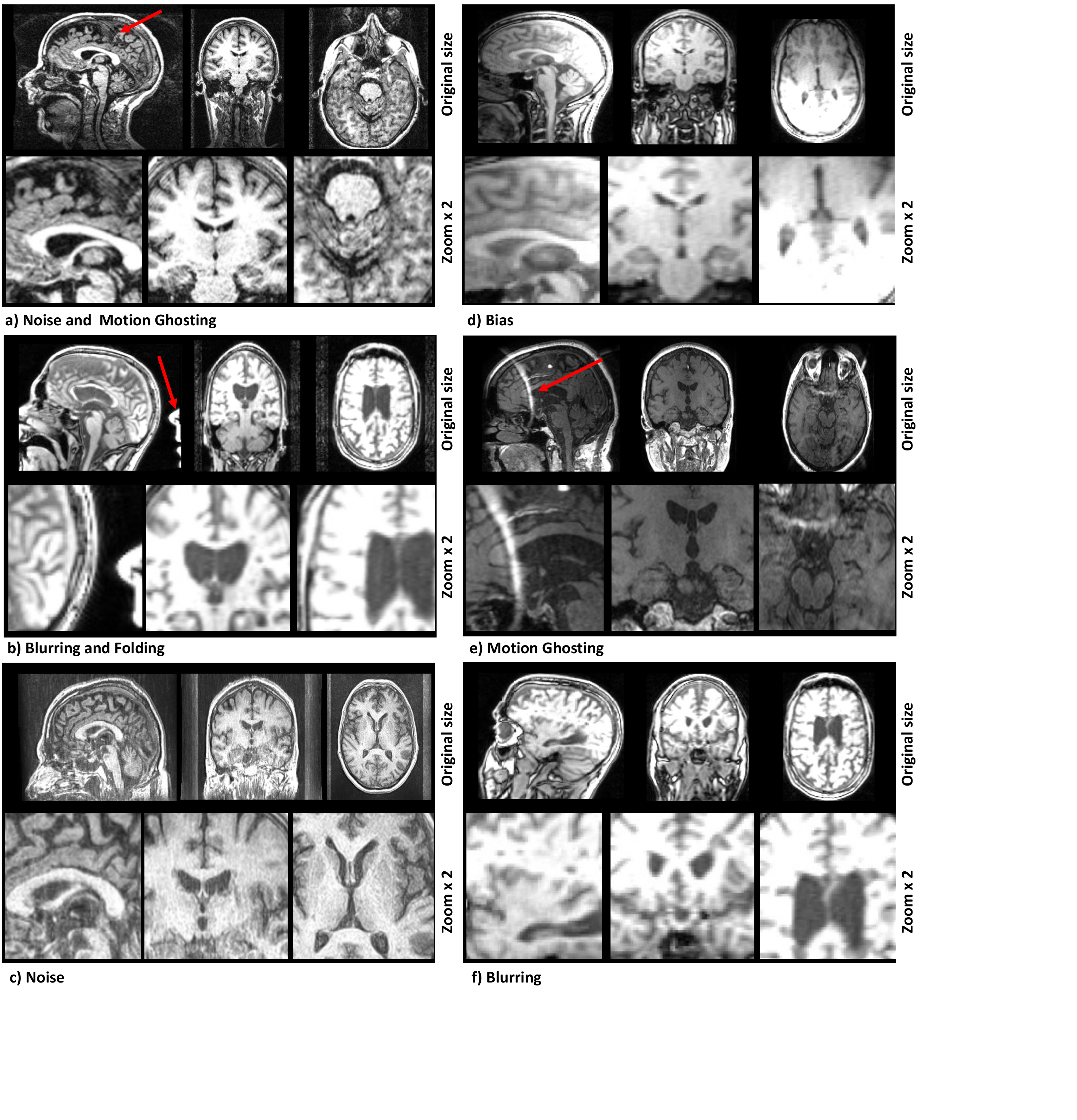}
\caption{Examples of scans from the real-world dataset having artefacts detected by our system  (true positive). Our approach detects scans with: noise (c), motion ghosting (a and e), blurring (b and f), folding (b) and bias (d).\label{fig:example}}
\end{center}
\end{figure*}

\begin{figure*}[t]
\begin{center}
\includegraphics[width=0.9\textwidth,trim={0cm 5cm 0cm 0cm},clip]{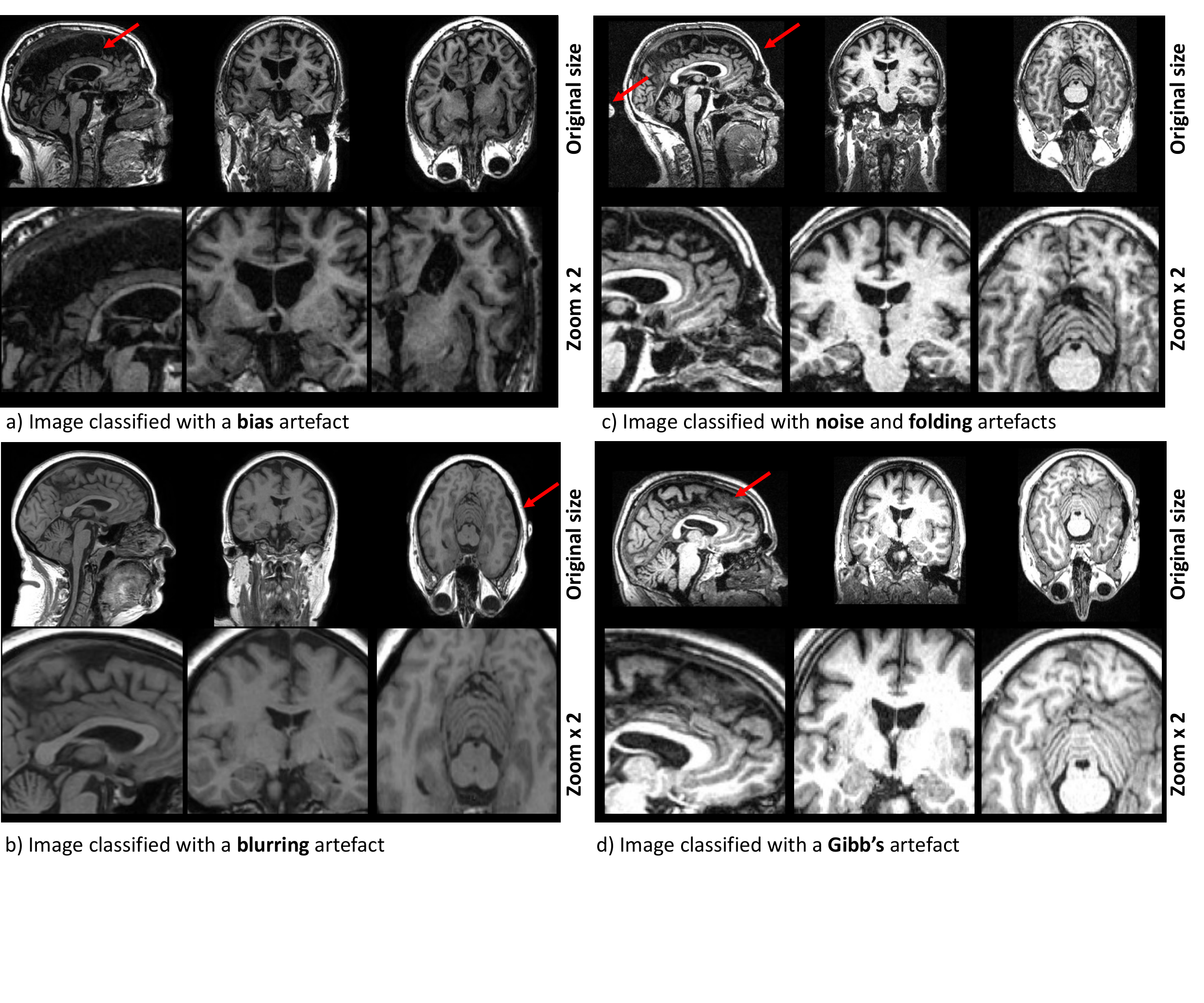}
\caption{Examples of misclassification obtained from our system on the real-world dataset. All these images were labelled as artefacts-free but detected by our model as having artefacts (false positive).\label{fig:example_bad}}
\end{center}
\end{figure*} 

\section{Experimental Results}
In our experiments, we first perform an ablation study (Section~\ref{sec:ablation}) where we test different configurations of our pipeline to assess the contributions of each proposed block. We then compare the proposed solution against state-of-the-art approaches on the two proposed datasets: i) the synthetic dataset $ADNI\_synt$ in Section~\ref{sec:comparison_synthetic}) and ii) the real-world MS dataset in Section~\ref{sec:comparison_real}. Finally, in Section~\ref{sec:computation_time} we assess the computation time for all the approaches to verify the real-time capability.
For the evaluation of the different approaches, we used 5 different metrics: accuracy, F1 score, F2 score, precision and recall. The F1 and F2 scores are defined as follows:

\begin{equation}
F1 = \frac{2 * Precision * Recall} { Precision + Recall} ,
\end{equation}
\begin{equation}
F2 = \frac{5 * Precision * Recall} {4* Precision + Recall} .    
\end{equation}

We believe that the combination of metrics that we have considered is adequate to validate our pipeline. In particular, the precision shows that the approach returns more relevant results than irrelevant ones and the recall that the algorithm returns most of the relevant results. F1 and F2 are instead a combination of precision and recall, where in the first case, there is a balanced weight on precision and recall and the latter less weight on precision and more weight on recall.

\subsection{Ablation study on $ADNI\_synt$}\label{sec:ablation}
Our ablation study is designed to analyze three components of the system: i) the feature extraction block $F$ (presented in Section~\ref{sec:features}), the feature selection $C$ (presented in  Section~\ref{sec:selection}) and the data augmentation obtained using the proposed artefact generators $S$ (presented in Section~\ref{sec:simulator}). Additionally, we proposed two different versions of our pipeline: i) a 2D version where only the slices from the centre of each MRI (first view) are used and ii) a 2.5D version where nine slices are instead extracted from $V$=3 different views (axial, sagittal and coronal) and $K$=3 different positions (1/3, 1/2, and 2/3 of the full size $s$).

The quantitative results obtained using our system with the test set from $ADNI\_synt$ are reported in Table~\ref{table:ablation_configuraion}. We can see that when no data augmentation is used during training (configuration $F$ and $F\_C$) the 2D version shows better performance than the 2.5D version. In particular, the configuration $F\_2.5D$ in comparison to its 2D counterpart ($F\_2D$) loses performance in the range of -1.8 and -4.2 percentage points across the different quality metrics, whereas the configuration $F\_C\_2.5D$ in comparison to the counterpart $F\_C\_2D$ loses performance in the range of 0 and -1.3 percentage points. This result was not expected and it is probably due to an overfitting problem during the training of the 2D version happening since the amount of data used in these configurations is limited and no data augmentation is used. On the other hand, when corrupted images are included during training (configurations $F\_S_2.5$ and $F\_C\_S$), this problem is overcome, and the 2.5D version provides better results than the 2D counterpart increasing the performances in the range of 0 and 1.2 percentage points. 

In Table~\ref{table:ablation_configuraion} we also assessed the contribution of each component of our system. The configuration with all the blocks enabled ($F\_C\_S$) is the one that has the highest performance (average performances are 97\%). The use of corrupted images (configurations $F\_S$) is the element that provides the largest improvement in our system (it increases performances in the range of +7.8\% and +9.5\% in comparison with the baseline) while the artefact-based feature selection process (configurations $F\_C$) increases performances in the range of +3.3\% and +5.4\% in comparison with the baseline. The combined effect of both these components (configurations $F\_C\_S$) increases performance in the range of +8.1\% and +10.7\%.

In Table~\ref{table:ablation_views}, we present the performance results of our system, obtained by experimenting with different numbers of slices and views. While we observed only minor improvements from changing the views, our results indicated that the axial view is optimal for identifying artefacts. Interestingly, increasing the number of slices from 1 to 3 was found to be more effective than changing the view. Ultimately, we can see that the highest performance was achieved when using 3 slices and all views (9 slices).

Finally, Table~\ref{table:ablation_featurees} provides a deeper understanding of how different features contribute to identifying various types of artefacts. This information is valuable in guiding future efforts to improve the accuracy of artefact detection systems by focusing on improving specific feature sets for certain types of artefacts.  From these results, we can see that the selection of different features for each artefact is based on their unique characteristics and underlying causes. For example, detecting incorrect labelling, motion ghosting, and smoothing can be challenging. Therefore, to accurately identify these artefacts, a comprehensive analysis of all available features is necessary. On the other hand, folding artefacts are primarily caused by violating the Nyquist criterion, which can be detected by analyzing the k-space data. Similarly, Gibbs artefacts are often due to undersampling or truncation in k-space, making them detectable by analyzing k-space features. Banding artefacts require examining both the k-space data and the image itself. In fact, in k-space, banding artefacts manifest as an outlier value, and in the image domain, they appear as alternating bright and dark bands, which can be quantified using metrics such as signal-to-noise ratio or contrast-to-noise ratio. Zipper artefacts manifest as a series of bright and dark lines in the image, and they can be easily characterized using features from the imaging domain. Finally, bias artefacts can result from various factors, such as uneven sensitivity profiles, shading, or calibration errors, requiring a combination of imaging and deep features to accurately identify the problem. In summary, it seems that the selection of different features for each artefact is based on the specific characteristics and underlying causes, and the optimal approach for detecting each artefact may vary accordingly.

\subsection{Comparison against related works on dataset containing artificially corrupted images}\label{sec:comparison_synthetic}
In this section, we present the results of the comparison of our approach against other methods. Obtained results are reported in Table~\ref{table:comparison} where we can see that our approach provides the highest performance in all the metrics. Notable is the comparison against the approach in~\citep{schlegl2019f}, which uses a generative model to learn the normal distribution (artefacts-free images) as an alternative solution to our solution, which instead learns to create artefacts directly. In general, our result shows that augmenting the training set with the proposed synthetic artefacts increases the performance of all approaches where it is applied. For example, we observed improvements between +12.8\% and +16.6\% percentage points on the approaches~\citep{schlegl2019f}+$\mathbf{S_b}$ (with no data augmentation) vs ~\citep{schlegl2019f}+$\mathbf{S_s}$ (with data augmentation) and improvements between +12.3 and +18.2 percentage points on the approaches~\citep{sadri2020mrqy}+$\mathbf{S_b}$(with no data augmentation) vs ~\citep{sadri2020mrqy}+$\mathbf{S_s}$(with data augmentation). 

\subsection{Comparison with related works on real-world data}\label{sec:comparison_real}
In this section, we present the results of comparing our approach against other methods using the proposed real-world MS dataset. The results of this experiment are reported in Table~\ref{table:comparisonARPEGGIO} and they show similar trends of improvements obtained from the synthetic dataset. In particular, the use of our data augmentation produces improvements between -0.5 and 1.6 percentage points on the approaches~\citep{schlegl2019f}+$\mathbf{S_b}$(with no data augmentation) vs ~\citep{schlegl2019f}+$\mathbf{S_s}$(with data augmentation) and improvements between +7.4 to +12.5 percentage points on the approaches~\citep{sadri2020mrqy}+$\mathbf{S_b}$(with no data augmentation) vs ~\citep{sadri2020mrqy}+$\mathbf{S_s}$(with data augmentation). Finally, the best configuration of our system improves performances between -0.8 and 7.3 percentage points in comparison with~\citep{schlegl2019f} and between 7.2 to 13.4 percentage points in comparison with~\citep{sadri2020mrqy}. Our proposed method resulted also in an improvement from 0.54 to 2.43, compared to a standard supervised framework~\cite{szegedy2016rethinking} trained with the same simulated artefacts used for data augmentation.

The statistical significance of the results obtained from our approach in comparison with the other approaches was assessed with a paired t-test where the p-values are all less than 0.0001 in both the synthetic dataset (Table~\ref{table:comparison}) and the real-world MS dataset (Table~\ref{table:comparisonARPEGGIO}). As for the ablation study (Table~\ref{table:ablation_configuraion}), we found that only the augmentation method involving simulated artefacts (S component) and the use of multiple slices (2.5D) when S is used demonstrated statistically significant improvements.

From this experiment, we also notice that all the different performances obtained on this dataset using our approach are still very high (in the range of [94\%-98\%]), confirming the capability of our approach to generalize well on the real-world dataset. This strong performance gives us two indications: the first is that our artificially corrupted images look sufficiently realistic and cover the variation in artefacts that may be expected in the real world, despite the fact that we test on an MS dataset and we based our training on artefacts added to data from Alzheimer's disease datasets (ADNI1 and ADNI2). The second is that our way to generate artefacts in 2D is a good approximation for generating artefacts in the entire 3D volume.

In Fig.~\ref{fig:example} we report some examples of images, from the real-world MS clinical dataset, correctly classified by our system (best configuration). In particular, we can see that our system is able to pick images demonstrating motion artefacts (a and e) (ghosting is clearly visible), scans demonstrating folding issues (b) (the nose is wrapped around), scans with noise (a and c), scans demonstrating signal bias (d), and finally, blurring (b and f) (the detail of the brain structures are here limited).

On the other hand, in Fig.~\ref{fig:example_bad} we also report some examples of misclassification obtained on the real-world MS clinical dataset. All these images were labelled by our experts as artefact-free but wrongly classified by our system (best configuration) as showing artefacts. In these cases, the images appear to have small artefacts or artefacts outside the brain that the experts have not considered relevant. This includes small bias imperfection (a), minimal smoothing (b), minimal folding and reduced noise (c), and limited Gibb's artefacts (d).

\subsection{Computational time and memory requirement}\label{sec:computation_time}
Table~\ref{table:computation_time} presents the computation time required to process a single scan using each of the approaches considered in our comparison for both feature extraction and classification. Feature extraction incurs the highest computational cost, taking between 0.27 and 0.81 seconds. On the other hand, final classification runs in the order of milliseconds with a relatively negligible computational cost. Although~\citep{schlegl2019f} employs only a small subset of features compared to our approach and provides the most efficient solution, our system yields superior performance, justifying a slightly higher computational cost (0.81 seconds instead of 0.27 seconds) that still achieves real-time processing.

Finally, the memory footprint for each component of our pipeline is as follows: $E$, $G$ and $D$ require altogether 9GB, the RES-NET requires 367MB, the block to extract imaging features $\xi(.)$ and the k-space features $\psi(.)$ require 20MB each and the final classifier $S_b$/$S_s$ 100MB each.

\section{Discussion and Conclusion}
In this work, we developed a semi-supervised approach to identify artefacts in brain MRI. Current state-of-the-art artefact classifiers in medical imaging have three key limitations: i) supervised approaches require a large set of data having labels at the pixel/voxel level that is time-consuming to be obtained, ii) unsupervised approaches trained to learn the distribution of artefacts-free images requires a large dataset of high-quality images that is hard to collect (images often have a small level of artefact), and iii) both supervised and unsupervised approaches often require high computation resources (i.e. voxel-level classification). 

To overcome these limitations we developed a new pipeline, which consists of i) a set of new physics-based artefact generators that are modelled and trained to learn to create artefacts, ii) a set of features from different domains extracted in real-time, iii) a feature selection block dependent on each class of artefact, and iv) a set of SVM classifiers. In particular, we used the artefact generators to augment our training set. Crucially, our artefact generation can be used to model artefacts that rarely occur. The experimental results show that augmenting the training set with corrupted scans substantially improves the classification performance.

We believe that in comparison with the state-of-the-art, our solution provides the best trade-off in terms of accuracy and processing time. Although we are only the second-best in time performance due to the large number of features used, our accuracy in detecting the images is higher while we still achieve real-time processing.

Finally, although we train our final classifiers in a supervised fashion, our solution has the advantage of using only artefact-free images with the benefit of requiring limited training labels (i.e. no pixel-based artefact delineation, no labels for each class of artefact). 

Our pipeline can be used to monitor the quality of MRI scans for research applications and in future may be used in clinical applications. Currently, quality assessment is carried out by human experts that verify when images yield good quality. However, with the increase in the amount of medical imaging data, manually identifying these artefacts is onerous and expensive. Automatic solutions are likely to replace human raters for large-scale repetitive tasks such as this one.

We see multiple further directions for future work. Firstly, our framework can be extended to model more artefacts (e.g. magnetic susceptibility, chemical shift, incomplete fat saturation, etc.). Additionally, we believe that our pipeline can be adapted beyond $T_{1}$-weighted MRI to other medical imaging contrasts, other organs and other imaging modalities, allowing us to have a comprehensive system with potential for future applications in research centres and hospitals.

To conclude, detecting artefacts in medical images presents a significant challenge due to the ambiguity of the artefact definition, which can be unclear even to experts and may vary depending on the specific clinical context. Similarly, our pipeline's training may have led to an overly sensitive model that detects good images as possible controversial artefacts, creating false positives. While false positives can be inconvenient and time-consuming to review, it is still preferable to have a system that is overly sensitive to anomalies and produces some false positives that can be screened out later by human experts than to have a system that misses actual anomalies or artefacts. Missing anomalies or artefacts can have serious consequences, particularly in medical settings where the accurate detection of abnormalities can impact patient diagnosis, treatment, and outcome, potentially leading to unnecessary procedures, delayed diagnoses, or even misdiagnosis, which can severely impact patient health.

Another limitation of our current solution is the lack of diversity in our dataset. Since all images used for training and creating artefacts come from ADNI, the generalization ability of our model may be limited. To address the issue of false positives and improve the generalization ability of our model, we plan to explore additional techniques such as  domain adaptation in future work. By incorporating domain-specific knowledge and adapting our model to the target domain, we hope to improve its ability to detect artefacts accurately and reduce the number of false positives.

\section*{Acknowledgements}

Data collection and sharing for this project was funded by the ADNI  (National Institutes of Health Grant U01 AG024904) and DOD ADNI (Department of Defense award number W81XWH-12-2-0012). ADNI is funded by the National Institute on Aging, the National Institute of Biomedical Imaging and Bioengineering, and through generous contributions from the following: AbbVie, Alzheimer's Association; Alzheimer's Drug Discovery Foundation; Araclon Biotech; BioClinica, Inc.; Biogen; Bristol-Myers Squibb Company; CereSpir, Inc.; Cogstate; Eisai Inc.; Elan Pharmaceuticals, Inc.; Eli Lilly and Company; EuroImmun; F. Hoffmann-La Roche Ltd and its affiliated company Genentech, Inc.; Fujirebio; GE Healthcare; IXICO Ltd.; Janssen Alzheimer Immunotherapy Research \& Development, LLC.; Johnson \& Johnson Pharmaceutical Research \& Development LLC.; Lumosity; Lundbeck; Merck \& Co., Inc.; Meso Scale Diagnostics, LLC.; NeuroRx Research; Neurotrack Technologies; Novartis Pharmaceuticals Corporation; Pfizer Inc.; Piramal Imaging; Servier; Takeda Pharmaceutical Company; and Transition Therapeutics. The Canadian Institutes of Health Research is providing funds to support ADNI clinical sites in Canada. Private sector contributions are facilitated by the Foundation for the National Institutes of Health (www.fnih.org). The grantee organization is the Northern California Institute for Research and Education, and the study is coordinated by the Alzheimer's Therapeutic Research Institute at the University of Southern California. ADNI data are disseminated by the Laboratory for Neuro Imaging at the University of Southern California. 

This project has received funding from Innovate UK - application number 74984.

FB, and DCA are supported by the NIHR biomedical research centre at UCLH.

\bibliographystyle{unsrtnat}
\bibliography{refs}  






\end{document}